\documentclass[journal,twocolumn,10pt]{IEEEtran}
\normalsize
\usepackage{amsmath,amssymb,amsfonts}
\usepackage{cite}
\usepackage{algorithmic}
\usepackage{algorithm}
\usepackage{alphalph,cite}
\usepackage{xcolor}
\usepackage{graphicx}
\usepackage{float}
\usepackage{stfloats}
\usepackage{textcomp}
\usepackage{multirow}
\usepackage{epstopdf}
\usepackage{subfig}
\usepackage{alphalph}
\usepackage{dirtytalk}
\usepackage{enumitem}
\usepackage{longtable}
\usepackage{comment}
\usepackage{stackengine}
\usepackage{pifont}
\usepackage{tikz}
\usepackage{caption}

\begin{document}
\bstctlcite{IEEEexample:BSTcontrol}
\title{Joint Spectrum Partitioning and Power Allocation for Energy Efficient Semi-Integrated Sensing and Communications}
\author{
	\IEEEauthorblockN{
		Ammar Mohamed Abouelmaati, Sylvester Aboagye, \textit{Member, IEEE}, and Hina Tabassum, \textit{Senior Member, IEEE}
\thanks{This work  was
supported by NSERC Discovery Grant funded by the Natural Sciences and
Engineering Research Council of Canada (NSERC). A. M. Abouelmaati and H. Tabassum are with the Department of Electrical Engineering and Computer Science at York University, Toronto, Canada; emails: ({ammarmo@my.yorku.ca, hinat@yorku.ca}).}
\thanks{S. Aboagye was with the Department of Electrical Engineering and Computer Science at York University, Toronto, Canada and  is currently with the School of Engineering, University of Guelph, Guelph, Canada. e-mail: (saboagye@uoguelph.ca).} 
}\vspace{-7mm}
}
\maketitle

\raggedbottom
\begin{abstract}
With spectrum resources becoming congested and the emergence of sensing-enabled wireless applications, conventional resource allocation methods need a revamp to support communications-only, sensing-only, and  integrated sensing and communication (ISaC) services together. In this letter, we propose two joint spectrum partitioning (SP) and power allocation (PA) schemes to maximize the aggregate sensing and communication performance  as well as corresponding energy efficiency (EE) of a semi-ISaC system that supports all three services in a unified manner. The proposed framework captures the priority of the distinct services, impact of target clutters,  power budget and bandwidth constraints, and sensing and communication quality-of-service (QoS)  requirements. We reveal that the former problem is jointly convex and the latter is a non-convex problem that can be solved optimally by exploiting fractional and parametric programming techniques. Numerical results verify the effectiveness of proposed schemes and extract novel insights related to the impact of the priority and QoS requirements of distinct services on the performance of semi-ISaC networks.  

\end{abstract}

\vspace*{-2mm}
\section{Introduction}


\textcolor{black}{Integrated Sensing and Communication (ISaC) enables simultaneous communication and sensing on network resources. However, using the entire spectrum for ISaC may be impractical due to existing allocations for various applications (e.g., L-band for air traffic control, S-band for weather observation). Also, not all users constantly require ISaC services. Hence, there is a need for a resource allocation framework supporting communication-only, sensing-only, and ISaC services referred to as \textit{semi-ISaC network}.}

Recently, there have been few studies that focused on resource allocation in ISaC systems.  In \cite{9945983}, an alternating optimization-based power and spectrum partitioning (SP) scheme was developed to optimize probability of detection (PD), Cramer-Rao bound (CRB), and the posterior CRB under constraints on the communication data rate. A power allocation (PA) scheme to maximize PD while satisfying minimum data rate was proposed in \cite{8378636}. The authors in \cite{9771575} considered communication-centric energy efficiency  (EE) maximization of an ISaC system by optimizing  beamforming vector under CRB constraint. In \cite{9976063}, the authors examined SP in an ISaC network to maximize the sum rate. However, no sensing performance metric was considered. Also,  \cite{9359480} focused on PA to maximize signal-to-interference-plus-noise ratio (SINR) for a radar and communication  coexistence scenario and \cite{10007634}  optimized beamforming vectors to enhance PD with constraints on the data rate requirements. 

\textcolor{black}{Unlike an ISAC system, the semi-ISAC system offers additional degree of freedom to network operators to dynamically prioritize users based on their required services (i.e., communication only, sensing only, or ISAC), minimum quality-of-service (QoS) requirements, and available network resources.}
Recently, in \cite{10036107}, outage probability analysis was performed for a single-input-single-output (SISO) semi-ISaC system and it was shown to outperform an ISaC system in terms of outage probability, ergodic rate, and radar estimation information rate. 

To date, none of the existing studies explored the joint optimization of SP and PA for ISaC systems, in general, and semi-ISaC systems, in particular.

\textcolor{black}{In this letter,  we develop a  unified sensing and communication optimization framework that jointly optimizes PA and SP for semi-ISAC networks under base-station (BS) power budget, total bandwidth, and users’ QoS  requirements while considering interference from clutters and users’ priorities. We consider two objectives of practical relevance, i.e., maximizing \textbf{(i)}  the aggregate radar mutual information (MI) and data rate, and \textbf{(ii)} EE. We establish the joint convexity of the former problem in spectrum and power allocation variables. The latter problem is a non-convex problem, thus we transform it into a concave-convex fractional problem and propose a Dinkelbach-based algorithm which converges to the global optimal solution. Simulation results affirm the optimized semi-ISaC framework's efficacy over existing benchmarks. Also, we illustrate how varying user requirements and priorities impact resource allocation choices.}


\begin{figure}
\centering
\includegraphics[width=1\linewidth,trim=4 4 4 4,clip]{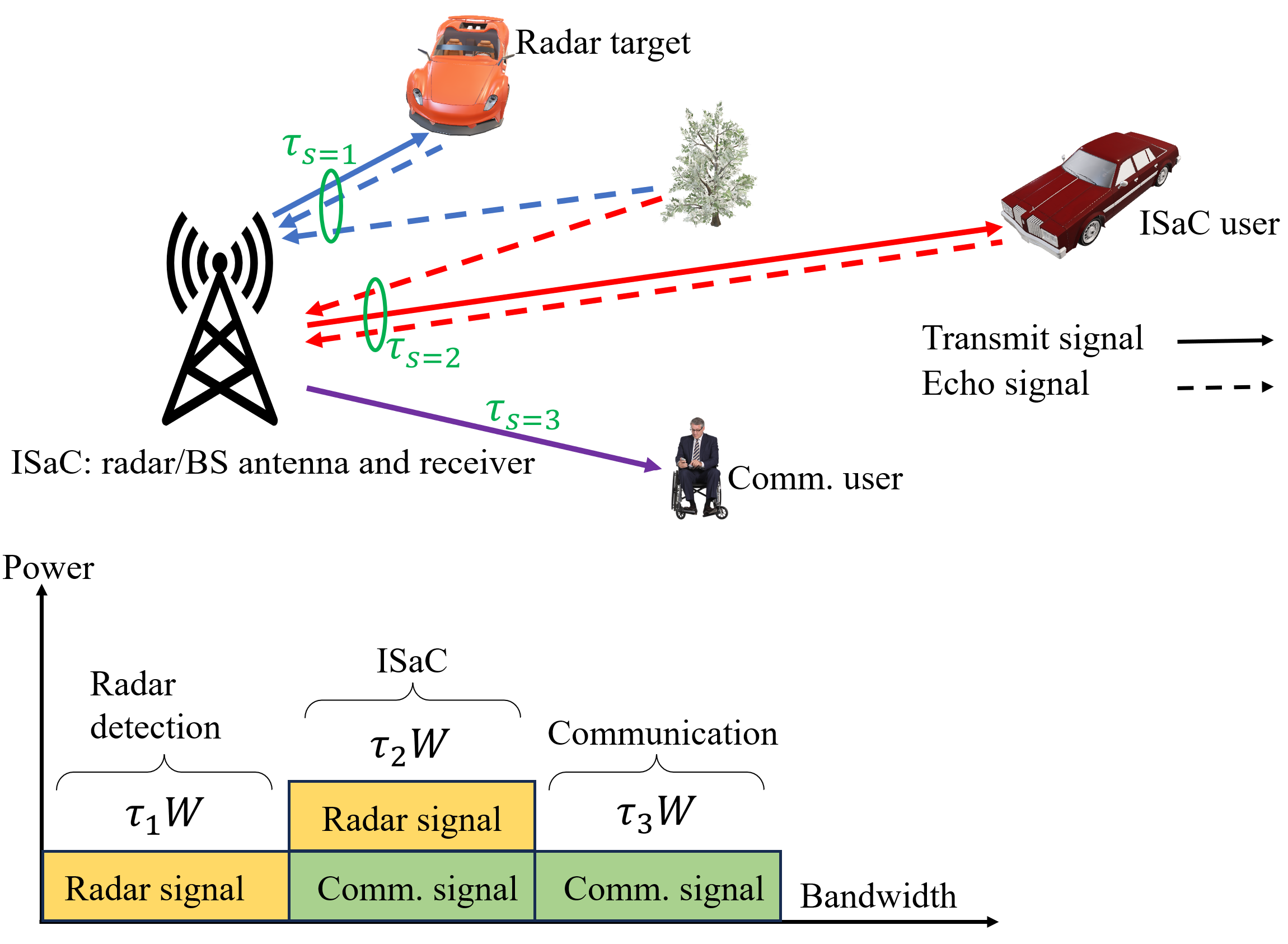}
      \caption{{Considered semi-ISaC system with sensing-only, ISaC, and communications-only services.}} 
       \label{sys_mod1}
\end{figure}

\vspace*{-2mm}
\section{System Model and Assumptions}
We consider the downlink (DL) of an orthogonal multiple access-based semi-ISaC system as illustrated in Fig.~\ref{sys_mod1}. The BS, operating in full-duplex mode,  partitions its available spectrum into three portions for three distinct services, namely, $\tau_1$ for sensing-only where BS senses the surrounding environment between it and user 1, $\tau_2$ for ISaC where BS senses the surrounding environment between it and user 2 while communicating with that user, and $\tau_3$ for communications-only with user 3. 
Note that the range resolution of the ISaC system is considered sufficiently accurate to avoid the interference between two targets (e.g., radar target and ISaC user) \cite{10036107}. Typical sensing applications include high-accuracy localization and tracking, augmented human sensing, simultaneous imaging and mapping, and space surveillance \cite{bayesteh2022integrated,8827589}. 

We define the proportion bandwidths $\tau_{s}$ such that the total bandwidth can be expressed as
$
W = {\tau_{1}W} + {\tau_{2}W} + {\tau_{3}W},
$
where  $\tau_{s} \in [0, 1]$ for $s = 1,2,3,$ and $\sum_{s=1}^3 \tau_{s} = 1$, as depicted in Fig.~\ref{sys_mod1}.   Without loss of generality, we assume  $J$ clutter scatterers observed at the radar receiver due to the
illumination of the radar transmitter. Similar to \cite{9828481,10036107,10007634}, we assume that the number of clutters and channel state information can be estimated perfectly. 

The path-loss model (PLM) on the communication channels is given as 
$
\mathcal{L}_c(d_s) =  \frac{ G_{\mathrm{tx}} d_s^{-\alpha_c} c^2}{(4 \pi f_c)^2} $\cite{8917703},
where $G_{\mathrm{tx}}$ is the transmit antenna gain, $s \in \{2,3\}$ is the index of the ISaC user and the communications-only user, $d_s$ denotes the link distance, $\alpha_c$ is the path-loss exponent (PLE), $f_c$ is the carrier frequency, and $c= 3 \times 10^8$ m/s is the speed of light.  
On the other hand, the two-way PLM for the radar echoes is given as
$
\mathcal{L}_r(d_1) = \frac{G_{\mathrm{tx}} d_1^{-2 \alpha_r  } \sigma_{\rm RCS} \lambda^2}{(4 \pi)^3}
$, where $\alpha_r$ is the PLE, $\lambda$ is the wavelength of the carrier, and $\sigma_{\rm RCS}$ 
is the target radar cross section (RCS). By replacing $d_1$ with $d_j$, the PLM can be customized for the clutter. In what follows, we describe the received signal models for sensing, ISaC, and communication scenarios.

\subsubsection{Sensing-Only} 
The  BS sends a radar signal to the target. The radar echo signal at the BS in the presence of $J$ clutter components can be expressed as \cite{9359480,10036107}: 
\begin{equation}\label{aug15a}y_1(t) =  {h_{1, d}h_{1, u} \sqrt{P_1\mathcal{L}_r(d_1)}z_1(t-\widetilde{\tau_1})} 
        + I_{\mathrm{c}_1} +  n_1(t),
    \end{equation}
where $I_{\mathrm{c}_1}= {\sum_{j=1}^J h_{j, d}h_{j, u} \sqrt{P_1 \mathcal{L}_r(d_{j})}z_1(t-\widetilde{\tau_j})}$ is the clutter interference, $P_1$ is the allocated transmit power, $z_1(t)$ is the radar  signal, $n_1(t)$ is the additive White Gaussian noise (AWGN) with variance $\sigma_1^2 = k_B T_{\rm temp} \tau_1 W$,  where $k_B$ is the Boltzmann constant,  $T_{\rm temp}$ is the absolute temperature, $\widetilde{\tau_1}$ and $\widetilde{\tau_j}$ are the time delay of the echo coming from user 1  and the $j$-th clutter, respectively.  It is assumed that the time delays can be perfectly compensated through synchronization  \cite{10036107}. The DL and uplink (UL)  small-scale fading channel power gains between the radar target and BS are given by $|h_{1,d}|^2$ and $|h_{1,u}|^2$, respectively.  The corresponding signal-to-clutter-plus-noise ratio (SCNR) is thus given by:
 
    \begin{equation}
    \gamma_1 = \frac{{P_{1} \mathcal{L}_r(d_1)\left|g_{1}\right|^2} }{{\sum_{j=1}^JP_{1} \mathcal{L}_r(d_{j}) \left|\zeta_{j}\right|^2}  + {\sigma_1^2} },\\
    \end{equation}
where $|g_{1}|^2 = |h_{1, d}|^2|h_{1, u}|^2$ is the cascaded channel power between the desired target and the BS,  and $|\zeta_{j}|^2 = |h_{j, d}|^2|h_{j, u}|^2$ is the cascaded  channel power of the $j$-th interfering clutter.

\subsubsection{ISaC} 
The framework for perceptive mobile networks--a joint communication and sensing enabled mobile network--is adopted here \cite{8827589}, where a single transmitted signal is used for communication and sensing, simultaneously. Specifically, the ISaC user utilizes the received signal for communication while downlink active sensing--where the BS uses reflected communication signals from its own transmitted signal for sensing--happens at the BS. Full-duplex technology allows the BS to perform sensing  and communication over the same time-frequency resources \cite{10158711}. Similar to \cite{8917703,9420261}, we assume perfect self-interference cancellation. The received signal at the ISaC user is given by: 
\begin{align}\label{isac2d}
        y_2^d(t) = {h_{2, d} \sqrt{P_{2} \mathcal{L}_c(d_2)}z_2(t)} + n_2(t),
\end{align}
and the corresponding SNR is expressed as:   
    \begin{equation}
        \gamma_{2}^d = \frac{{P_2 \mathcal{L}_c(d_2) |h_{2, d}|^2}}{ {\sigma_2^2}},
    \end{equation}
where $P_2$ is the transmit power of the ISaC signal $z_2{(t)}$,  $n_2(t)$ is AWGN with variance $\sigma_2^2 = k_B T_{\rm temp} \tau_2 W$, and $|h_{2,d}|^2$ is the channel power. The reflected echo at the BS in the presence of $J$ clutter components can be expressed as:
\begin{multline}\label{isac2u}
        y_2^u(t) = {h_{2, d}h_{2, u} \sqrt{P_2 \mathcal{L}_r(d_2)}  z_2(t-\widetilde{\tau_2})}+ I_{\mathrm{c}_2} +n_2(t),
    \end{multline}   
   where $I_{\mathrm{c}_2}=
         {\sum_{j=1}^J h_{j, d}h_{j, u} \sqrt{P_{2} \mathcal{L}_r(d_{j})}z_2(t-\widetilde{\tau_j})}$, and the corresponding SCNR is given by:
          
    \begin{equation}
        \gamma_{2}^u = \frac{{P_2 \mathcal{L}_r(d_2) |g_{2}|^2}}{ \displaystyle \sum_{j=1}^J {P_{2} \mathcal{L}_r(d_j) \left|\zeta_{j}\right|^2}+  {\sigma_2^2}},
    \end{equation}
where $|g_{2}|^2 = |h_{2, d}|^2|h_{2, u}|^2$ is the cascaded channel power gain.

 \subsubsection{Communication Signal} The received signal at the communications-only user is given by:
 \begin{equation}\label{dr}
        y_3(t) = {h_{3} \sqrt{P_{3} \mathcal{L}_c(d_3)}z_3(t)}  + n_3(t),
    \end{equation}
and the corresponding SNR is 
$
    \gamma_3 = \frac{P_3 \mathcal{L}_c(d_3) \left|h_3\right|^2}{\sigma_3^2},
$
where $z_3{(t)}$ is the transmit signal, $|h_3|^2$ is the small-scale fading channel power gain, $n_3(t)$ is AWGN with variance $\sigma_3^2 = k_B T_{\rm temp} \tau_3 W$, and $P_3$ is its allocated transmit power.

\section{MI, Data Rate, and EE Optimization}
\subsection{MI and Data Rate Optimization}
  For the radar sensing and ISaC transmission, we use MI as a   metric that measures how much information about the channel is conveyed to the radar receiver \cite{9528013,8889694}. Specifically, after receiving $y(t)$, the priori uncertainty of the target/ISaC user decreases since there is some information about the target/ISaC user contained in the cascaded channel gain $g$. The MI between $y(t)$ and $g$ is given as $I(y;g)=H(y)-H(y|g)=H(y)-H(n_o)$ \cite{9528013,8889694}, where $n_o$ is the noise and interference at the receiver. {\textcolor{black}{Assuming both $H(y)$ and $H(n_o)$ are Gaussian, their differential entropy is given as $H(y)=\frac{1}{2}\log_2\left(2\pi e E \{|y|^2  \}  \right)$ and $H(n_o)=\frac{1}{2}\log_2\left(2\pi e E \{|n_o|^2  \}  \right)$ \cite{8889694,Cover-Thomas2012}. The MI can be given as}}:
\begin{align}
       I=H(y)-H(n_o)=\frac{1}{2}\log_2\left(E  \left\{\frac{|y|^2}{|n_o|^2}  \right\}  \right)
\end{align}
For radar sensing, the MI at the BS according to (\ref{aug15a}) can be given as $ I_{1}=\tau_1 W \log_2(1 + \gamma_1)$.
{\textcolor{black}{Based on (\ref{isac2d}) and (\ref{isac2u}), the MI at the ISaC user (i.e., communication performance) and at the  BS (i.e., sensing performance) can be given as $I_{2}^d=\tau_2 W \log_2(1 + \gamma_2^d)$ and $I_{2}^u=\tau_2 W \log_2(1 + \gamma_2^u)$, respectively. Thus, $I_2=I_{2}^d + I_{2}^u$. 
}}
For communications, according to (\ref{dr}), the data rate can be determined by $I_3 = \tau_3 W  \log_2 (1 + \gamma_3)$.

The unified optimization framework to maximize the weighted sum of the MI and data rate via the joint optimization of SP and PA can be mathematically formulated as follows:
\begin{equation}\label{eepp}
\centering
\begin{array}{ll}
\mathop {\max }\limits_{{\bf P},{\boldsymbol \tau}}\displaystyle \sum_{s=1}^{3} \Gamma_s \tau_s \mathrm{log}_2(1 + \gamma_s)  & {\rm s.t.} \\
\mathbf{C}_1: \tau_1 W \log_2(1 + \gamma_1)\ge R_{\rm r}, & \mathbf{C}_5:\sum\limits_{s=1}^{3} \tau_{s} =1,\\
\mathbf{C}_2: \tau_2 W \log_2(1 + \gamma_2^d)\ge R_{\rm c}, & \mathbf{C}_6:\sum\limits_{s=1}^{3} P_{s} \le P_{\max},  \\
\mathbf{C}_3: \tau_2 W \log_2(1 + \gamma_2^u)\ge R_{\rm r}, & \mathbf{C}_7:\tau_{s},P_s > 0,\,\forall s, \\
\mathbf{C}_4: \tau_3 W  \log_2 (1 + \gamma_3) \ge R_{\rm c}, & \\
\end{array}
\end{equation}
where $\Gamma_s\ge 0$ is the priority of the $s$-th user, $R_r$ and $R_c$ are the minimum sensing and communication QoS requirements, and $P_{\max}$ is the maximum allowable transmit power. The constraints $\mathbf{C}_1$, $\mathbf{C}_2$, $\mathbf{C}_3$, and $\mathbf{C}_4$ are the QoS requirements for radar target, ISaC user's downlink communication, ISaC user's uplink sensing,  and communication user, respectively. 
Constraint $\mathbf{C}_5$ ensures that the sum of the bandwidth coefficients must be 1. Constraint $\mathbf{C}_6$ is the transmit power budget. Constraint $\mathbf{C}_7$ ensures that the SP and PA variables are non-negative. 

It is shown in the sequel that the optimization problem in (\ref{eepp}) is convex. Let ${\mathcal M}(x, q) = Wx \log_2 (1 + \frac{a_sq}{b_sq + c_sx})$, \textbf{dom} ${\mathcal M} = \{x, q | 0 < x \le 1, q > 0 \}$, where $a_s$, $b_s$, and $c_s$ are non-negative constants. ${\mathcal M}(x, q)$ represents the MI expression with clutter interference for the radar target ($s=1$) and ISaC user ($s=2$). For $s=1$, $a_1 =  G_{\mathrm{tx}} d_1^{-2\alpha_r} \sigma_{\rm RCS} \lambda^2\left|g_{1}\right|^2$, $b_1 = \sum_{j=1}^J G_{\mathrm{tx}} d_j^{-2\alpha_r} \sigma_{\rm RCS} \lambda^2 \left|\zeta_{j}\right|^2$ and $c_1 = (4 \pi)^3k_B T_{\rm temp} W$. 
For $s=2$, $a_2 = G_{\mathrm{tx}} d_2^{-2\alpha_r} \sigma_{\rm RCS} \lambda^2 \left|g_{2}\right|^2$, $b_2 = \sum_{j=1}^J G_{\mathrm{tx}} d_j^{-2\alpha_r} \sigma_{\rm RCS} \lambda^2 \left|\zeta_{j}\right|^2$, and $c_2 = (4 \pi)^3k_B T_{\mathrm{temp}} W$.

Also, ${\mathcal K}(x, q) = Wx \log_2 (1 + \frac{dq}{ex})$ \textbf{dom} ${\mathcal K} = \{x | 0 < x \le 1, q>0\}$, where $d, e >0$,  denotes the  data rate of the communication  user ($s=3$). 
$d$ is  $d =  G_{\mathrm{tx}} d_3^{-\alpha_c}  c^2\left|h_{3}\right|^2$, and $e$ is given by $e = (4 \pi f_c)^2 k_B T_{\mathrm{temp}}W$.

We show that the multivariable functions ${\mathcal M}$ and ${\mathcal K}$ are concave by analyzing their Hessian matrices. 
The Hessian matrix for  ${\mathcal M}(x,q)$ is given by (\ref{mHess}). 
\begin{figure*}
\begin{equation} \label{mHess}
   H_{\mathcal M} =  \left[\begin{matrix}
     -\frac{W(a^2xq^2c^2+2axq^2bc^2+2aq^3b^2c+2a^2q^3bc)}{\ln \left(2\right)\left(xc+aq+qb\right)^2\left(xc+qb\right)^2} \\ \\
 \frac{W(2axq^2b^2c+2a^2xq^2bc+a^2x^2qc^2+2ax^2qbc^2)}{\ln \left(2\right)\left(xc+aq+qb\right)^2\left(xc+qb\right)^2}
    \end{matrix} \quad\begin{matrix} \frac{acW\left(2x^2qbc+ax^2qc+2xq^2b^2+2axq^2b\right)}{\ln \left(2\right)\left(xc+aq+qb\right)^2\left(xc+qb\right)^2}\\ \\
-\frac{ax^2cW\left(2qb^2+2aqb+axc+2xbc\right)}{\ln \left(2\right)\left(xc+aq+qb\right)^2\left(xc+qb\right)^2}
\end{matrix}\right].
\end{equation}
\hrule
\vspace*{-3mm}
\end{figure*}
Clearly, the determinant of the first leading principle $D_1$, denoted  $|D_1|$, is less than zero. The determinant of the second leading principle $|D_2|=|H_{\mathcal M}| = 0$. This concludes that the MI function is concave.

Similarly for ${\mathcal K}(x,q)$, the Hessian matrix is given by
\begin{equation}
H_{\mathcal K} = 
\begin{bmatrix}
-\frac{a^2q^2}{\ln \left(2\right)x\left(xb+aq\right)^2} & \frac{a^2q}{\ln \left(2\right)\left(xb+aq\right)^2} \\
\frac{a^2q}{\ln \left(2\right)\left(xb+aq\right)^2} & -\frac{a^2x}{\ln \left(2\right)\left(xb+aq\right)^2} 
\end{bmatrix}.
\end{equation}
For this matrix, it can be shown that $|D_1|<0$ and $|D_2| = |H_{\mathcal K}| =0$. This concludes that the data rate expression is concave. Since the sum of concave functions is concave and the constraint sets are either concave or affine, we can conclude that the optimization problem in (\ref{eepp}) is a convex problem which can be  solved using standard convex optimization solvers.

\vspace*{-3mm}
\subsection{Energy Efficiency Optimization}
EE in semi-ISaC systems can be defined as the ratio of the sum of the data rate and MI to the total power consumed. Mathematically, it can be written as:

\begin{equation}\label{EE_def}
    EE = \frac{{\mathcal A}({\bf P}, {\boldsymbol{\tau}})} {{\mathcal B}({\bf P})}=\frac{\sum_{s=1}^{3} \Gamma_s \tau_s \mathrm{log}_2(1 + \gamma_s)}{P_1 + P_2 + P_3 + \omega}
\end{equation}
where $\omega$ denotes the circuit power consumption. The unified joint SP and PA to maximize EE in semi-ISaC networks can be formulated as:
\begin{equation}\label{eepp1}
\centering
\begin{array}{l}
\mathop {\max }\limits_{{\bf P},{\boldsymbol \tau}} \,\, EE  \\
{\rm{s}}{\rm{.t}}{\rm{.}}\,\,\,
\mathbf{C}_1,
\mathbf{C}_2,
\mathbf{C}_3,
\mathbf{C}_4,
\mathbf{C}_5,
\mathbf{C}_6,
\mathbf{C}_7.
\end{array}
\end{equation}
\begin{algorithm}[t]
\caption{Optimal SP and PA in Semi-ISaC Algorithm}
 \label{algo1}
 \begin{algorithmic}[1]
\STATE {set $j=0$, $\delta >0, \eta)j = 0, F(\eta) > 0$;}
\STATE{{\textbf{while} $F(\eta_j) > \delta$} do}
\STATE {Solve (\ref{eepp3}) to obtain $P_j$ and $\tau_j$;}
\STATE {Update $j \leftarrow j + 1$;}
\STATE {Determine $\eta_j \leftarrow {{\mathcal A}({P_j}, {{\tau_j}})}/ {{\mathcal B}({P_j}, {{\tau_j}})}$;}
\STATE{Set $F(\eta_j)$ to objective value of (\ref{eepp3})}
\STATE {{\textbf{end while}}}
\STATE {{\textbf{output:} ${\bf P}^*, {\boldsymbol{\tau}}^*$}}
 \end{algorithmic}
\end{algorithm}
The optimization problem in (\ref{eepp1}) is a concave-convex fractional program since the the numerator of the objective function is concave and the denominator is affine. By introducing the parameter $\eta$, the fractional objective function in  (\ref{eepp1}) can be transformed to an equivalent non-fractional form, and the EE optimization for a given $\eta$ can be reformulated as:
\begin{equation}\label{eepp3}
\centering
\begin{array}{l}
    F(\eta) = \mathop {\max }\limits_{{\bf P},{\boldsymbol \tau}} {{\mathcal A}({\bf P}, {\boldsymbol{\tau}})} - \eta {{\mathcal B}({\bf P})} \\
    {\rm{s}}{\rm{.t}}{\rm{.}}\,\,
    \mathbf{C}_1,
    \mathbf{C}_2,
    \mathbf{C}_3,
    \mathbf{C}_4,
    \mathbf{C}_5,
    \mathbf{C}_6,
    \mathbf{C}_7. 
\end{array}
\end{equation}
The optimal solution ${\bf P}^*$ and ${\boldsymbol \tau}^*$ to (\ref{eepp1}) is also optimal for (\ref{eepp3}) for a certain $\eta^* \ge 0$ that satisfies $F(\eta^*) = 0$. For a given $\eta$, problem (\ref{eepp3}) is a convex problem in terms of ${\bf P}$ and ${\boldsymbol \tau}$, therefore we employ an iterative Dinkelbach-type algorithm to compute the optimal PA and SP by updating $\eta$ via (\ref{EE_def}) to find the roots of the equation $F(\eta) = 0$. The proposed algorithm is summarized in Algorithm~\ref{algo1}. \textcolor{black}{The EE problem is in fact a multi-objective optimization problem for different values of $\eta$}.

The Dinkelbach method has been shown to have a linear convergence rate \cite{crouzeix1985algorithm}. In addition, it can be seen that since ${{\mathcal A}({\bf P}, {\boldsymbol{\tau}})}$ and ${{\mathcal B}({\bf P})}$ are  concave and convex, respectively, and the constraints form a convex set, Algorithm~1 requires to solve a convex problem in each iteration, which can be accomplished with polynomial complexity.

\begin{figure*}
\centering
 \begin{minipage}{0.32\textwidth}
 \centering
  \includegraphics[width=1\linewidth]{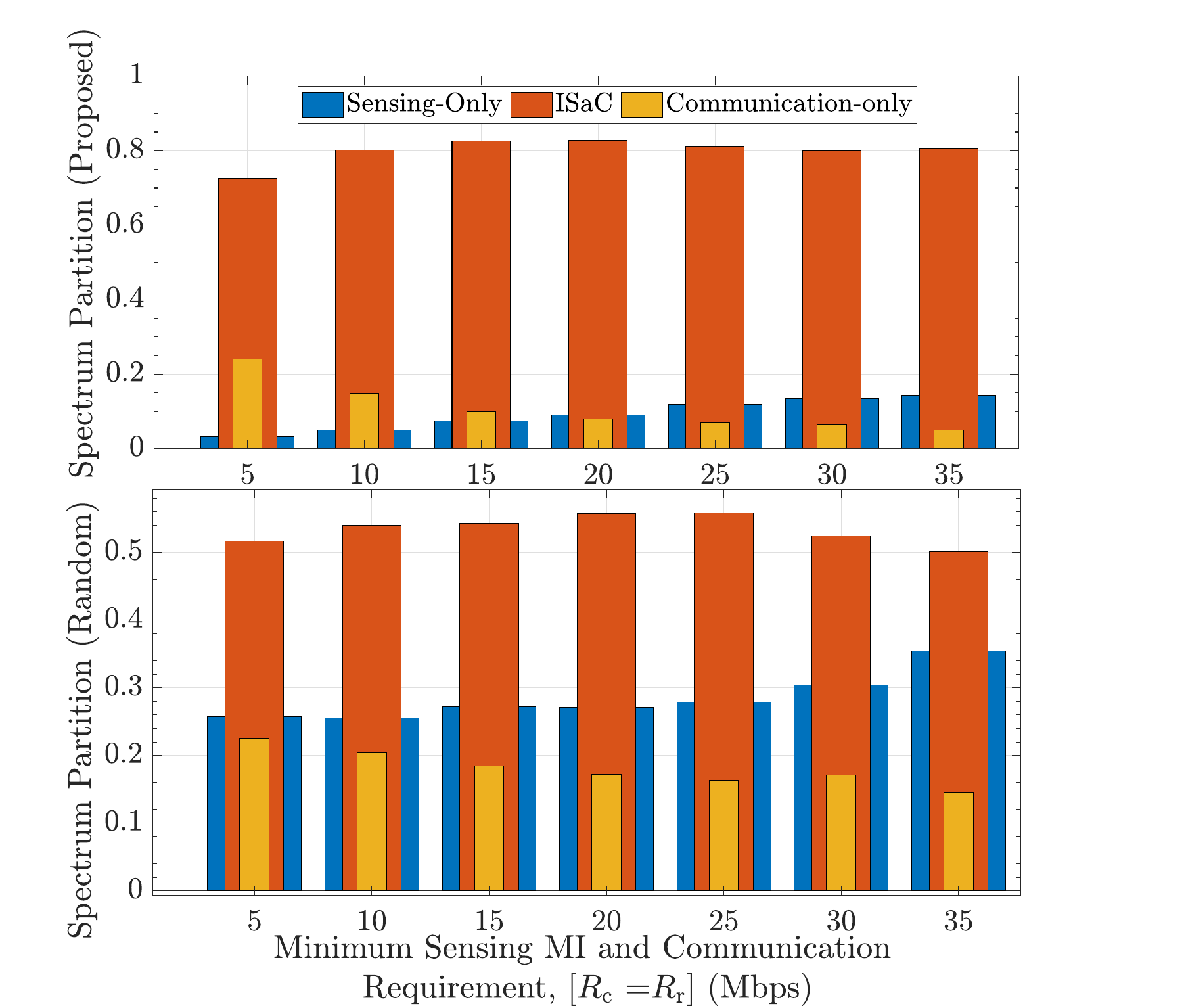}
      \caption{{{Spectrum Partition as a function of sensing MI threshold ($R_r$) and data rate threshold ($R_c$) when $R_c=R_r$.}}}
       \label{S_part}
\end{minipage}\hfill
\begin{minipage}{0.32\textwidth}
 \centering
\includegraphics[width=1\linewidth]{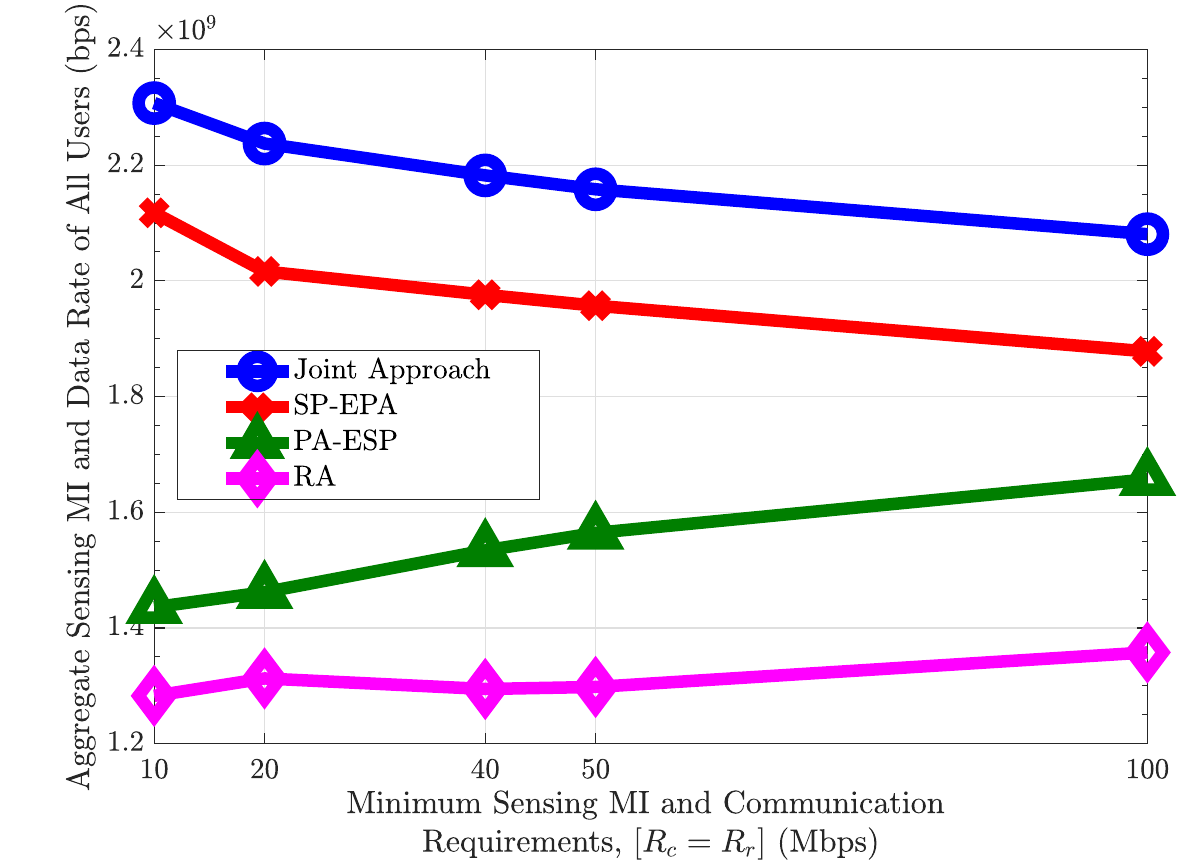}
      \caption{{{Aggregate sensing MI and data rate as a function of sensing MI threshold ($R_r$) and data rate threshold ($R_c$).}}}
       \label{MI_QoS}
\end{minipage}\hfill
 \begin{minipage}{0.32\textwidth}
 \centering
 \includegraphics[width=1\linewidth]{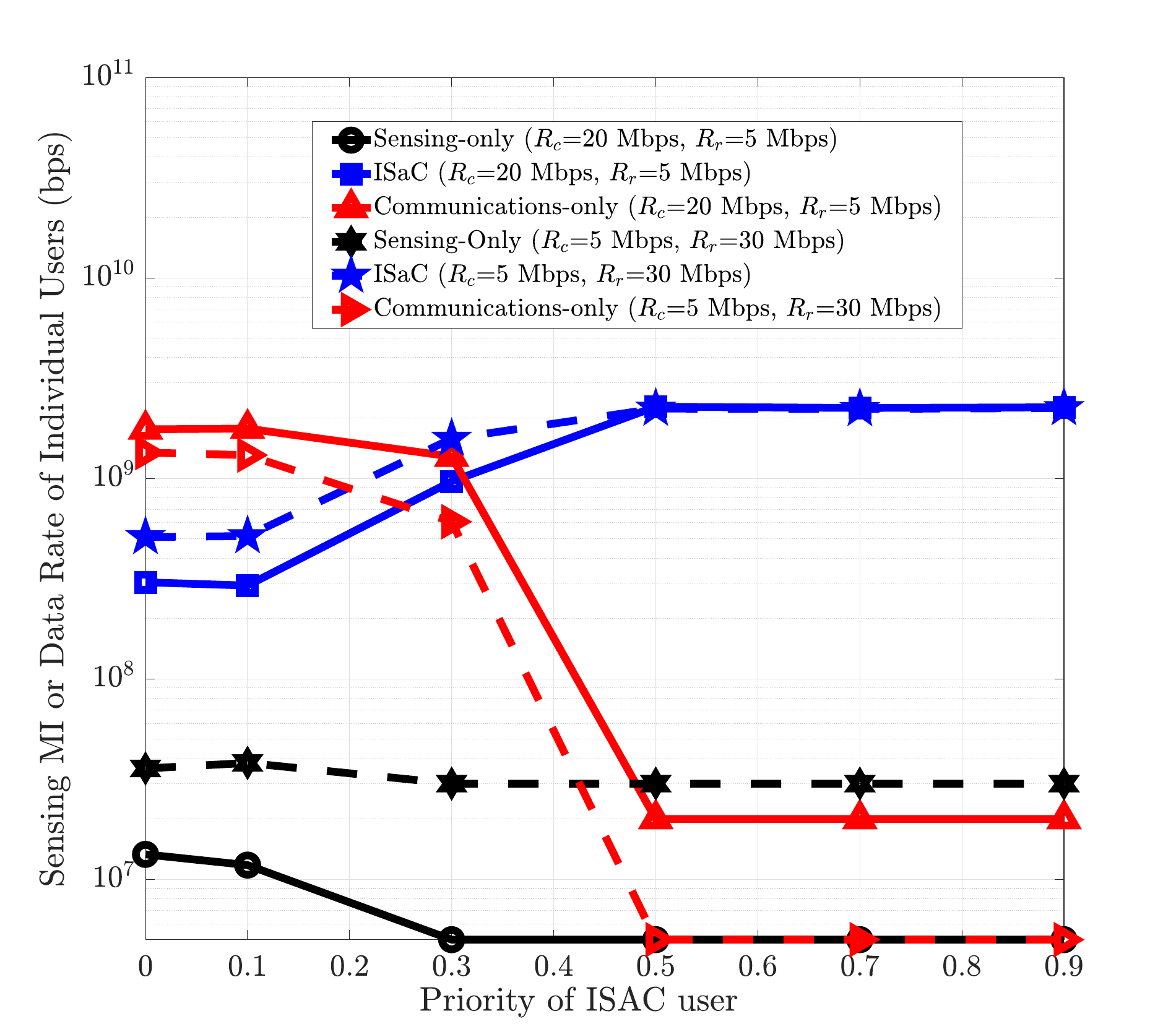}
      \caption{{{Sensing MI or data rate of each user as a function of the priority of ISaC user and minimum QoS requirements.}}}
       \label{MI_gamma3}
\end{minipage}%
\vspace*{-2mm}
\end{figure*}

\vspace*{-2mm}
\section{Simulation results and Discussions}
In this section, we perform comprehensive simulations to evaluate the effectiveness of the proposed algorithms. A cell radius of 40 m is considered with uniformly distributed  users. Unless stated otherwise, the following simulation parameters have been used, i.e., $W=100$ MHz \cite{10007634},  $T_{\rm temp} = 724$ K, $f_c = 10$ GHz \cite{10036107}, $\sigma_{\rm RCS}= 0.1$ {\color{black} $m^2$}\cite{10036107}, $\alpha_r = \alpha_c = 2.5$ \cite{10036107,8917703}, $P_{\max}= 46$ dBm,  $J=2$ \cite{10007634}, $|\zeta_{1}|^2 = 0.01, |\zeta_2|^2=0.001$  \cite{10007634}, {\color{black}$\omega = 33$ dBm \cite{9771575}}, $R_r = 5$ Mbps, and $R_c = 20$ Mbps. The users' priorities are set as $\Gamma_1=\Gamma_2=\Gamma_3=1/3$ \cite{9828481}. The small-scale fading is modeled according to Nakagami-$m$ fading, with $m=3$. The following benchmarks are used for performance comparison. SP with equal power allocation (SP-EPA), PA with equal spectrum partitioning (PA-ESP),  and random allocation (RA) with feasible solutions.

Fig.~\ref{S_part} depicts that the portion of the spectrum that is allocated to the different users vary significantly between the proposed approach and the RA scheme. For example, less spectrum is allocated to the ISaC user in RA scheme which underscores the significance of allocating more spectrum to ISaC user for optimal performance. The spectrum allocations for SP-EPA and PA-ESP are not shown since the former follows the trend for the proposed scheme and the latter has no spectrum optimization involved.

Figure~\ref{MI_QoS} depicts the impact of varying the minimum communication and sensing QoS requirements on the sum of MI and data rate considering the proposed approach and benchmarks. It can be observed that the proposed joint approach outperforms the baseline schemes by achieving an average of 10$\%$, 43$\%$, and 67$\%$ performance gain as compared to the SP-EPA, PA-ESP, and RA schemes, respectively. This highlights the significance of joint SP and PA. Moreover, while the performance of the joint approach and the SP-EPA  reduces with increasing $R_c$ and $R_r$, the sum of MI and data rate for the PA-ESP and RA benchmarks slightly improves. These trade-offs can be explained as follows. Typically, the communication-only user contributes more to the objective function (i.e., aggregate sensing MI and data rate) due to one-way PL effect. Hence, more spectrum is allocated to communication-only user as compared to sensing users (as shown in Fig.~\ref{S_part}). Subsequently, the sensing  performance almost always achieves $R_r$ at most, whereas communication user almost always exceeds $R_c$.
When $R_c$ and $R_r$ increase, more spectrum is needed to guarantee the higher sensing QoS requirement due to the two-way path-loss. Thus, the spectrum allocated to communication-only user reduces while that of sensing-only and ISaC users increases, as evident in Fig.~\ref{S_part}. This spectrum reduction leads to the reduction in the aggregate sensing MI and data rate of all users for the proposed joint approach and the SP-EPA method as shown in Fig.~3. Note that the performance of the PA-ESP and RA improves with increasing $R_c$ and $R_r$ as no SP optimization is involved.

Figure~\ref{MI_gamma3} examines the effect of  varying the priority of ISaC user $\Gamma_2$ on the performance of the semi-ISaC network for different QoS threshold values. In this figure, the priority of the ISaC user $\Gamma_2$ is varied by fixing $\Gamma_1=\Gamma_3$ such that $\Gamma_1=\Gamma_3=(1-\Gamma_2)/2$. As expected,  the performance of the communication-only user and the  sensing-only user decreases, while  the MI for the ISaC user improves.  However, the performance loss of sensing is evident when sensing requirement $R_r$ is small and the  loss is mostly insignificant compared to communications-only user. Moreover, while setting $R_r > R_c$ reduces the data rate of the communication-only user, the MI of the sensing-only user and ISaC user increases. This is due to the fact that increasing $R_r$ from 5 Mbps to 30 Mbps while reducing $R_c$ from 20 Mbps to 5 Mbps means allocating more resources to satisfy higher sensing QoS requirement. For $\Gamma_2 \ge 0.5$, the curves when $R_r < R_c$ and $R_r > R_c$ for the ISaC user almost overlap since the user has very high priority.

\begin{figure*}
\centering
 \begin{minipage}{0.32\textwidth}
 \centering
 \includegraphics[width=1\linewidth]{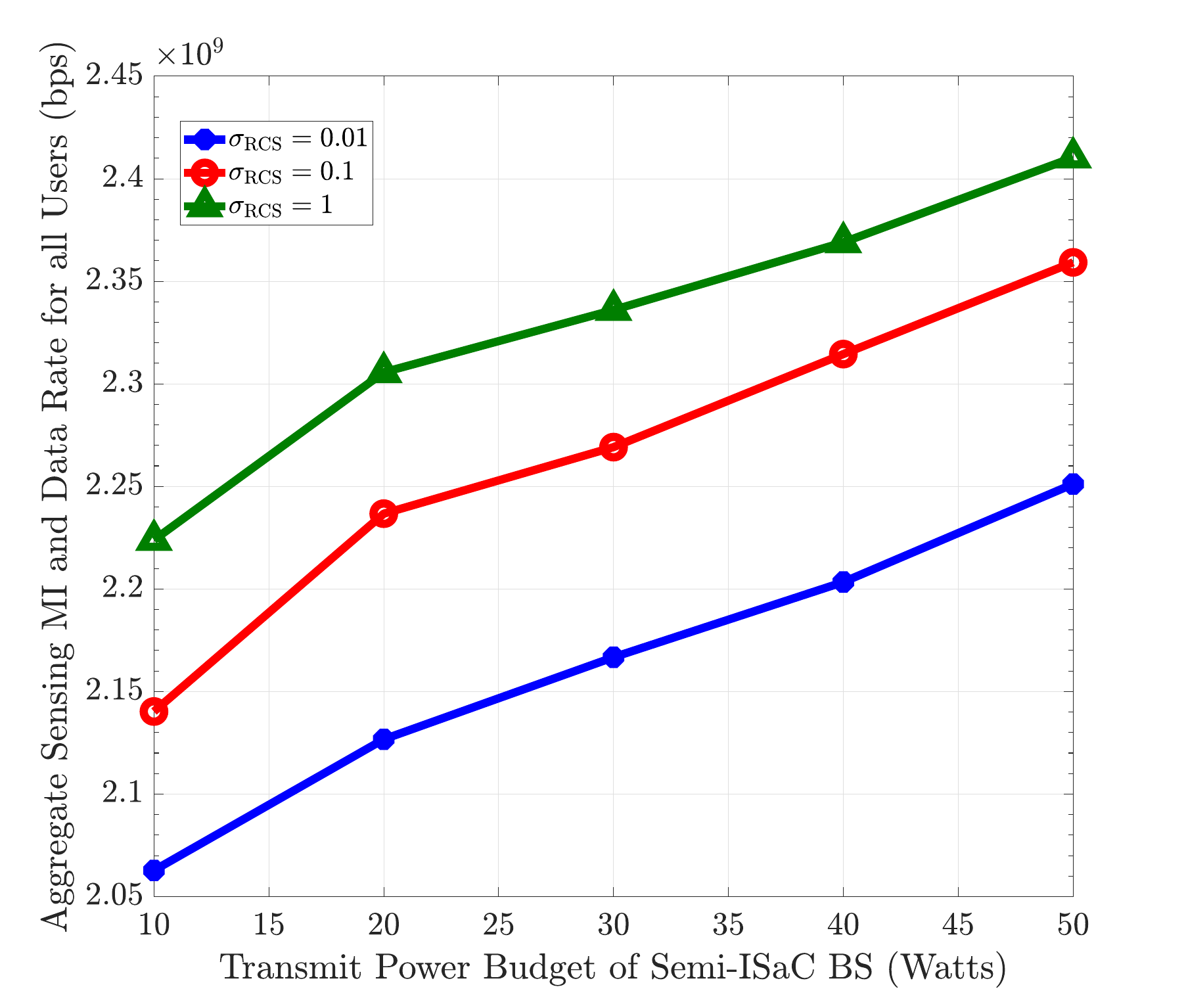}
      \caption{{{Aggregate sensing MI and data rate as a function of transmit power of Semi-ISaC BS and target RCS.}}}
       \label{MI_rcs}  
\end{minipage}\hfill
\begin{minipage}{0.32\textwidth}
 \centering
 \includegraphics[width=1\linewidth]{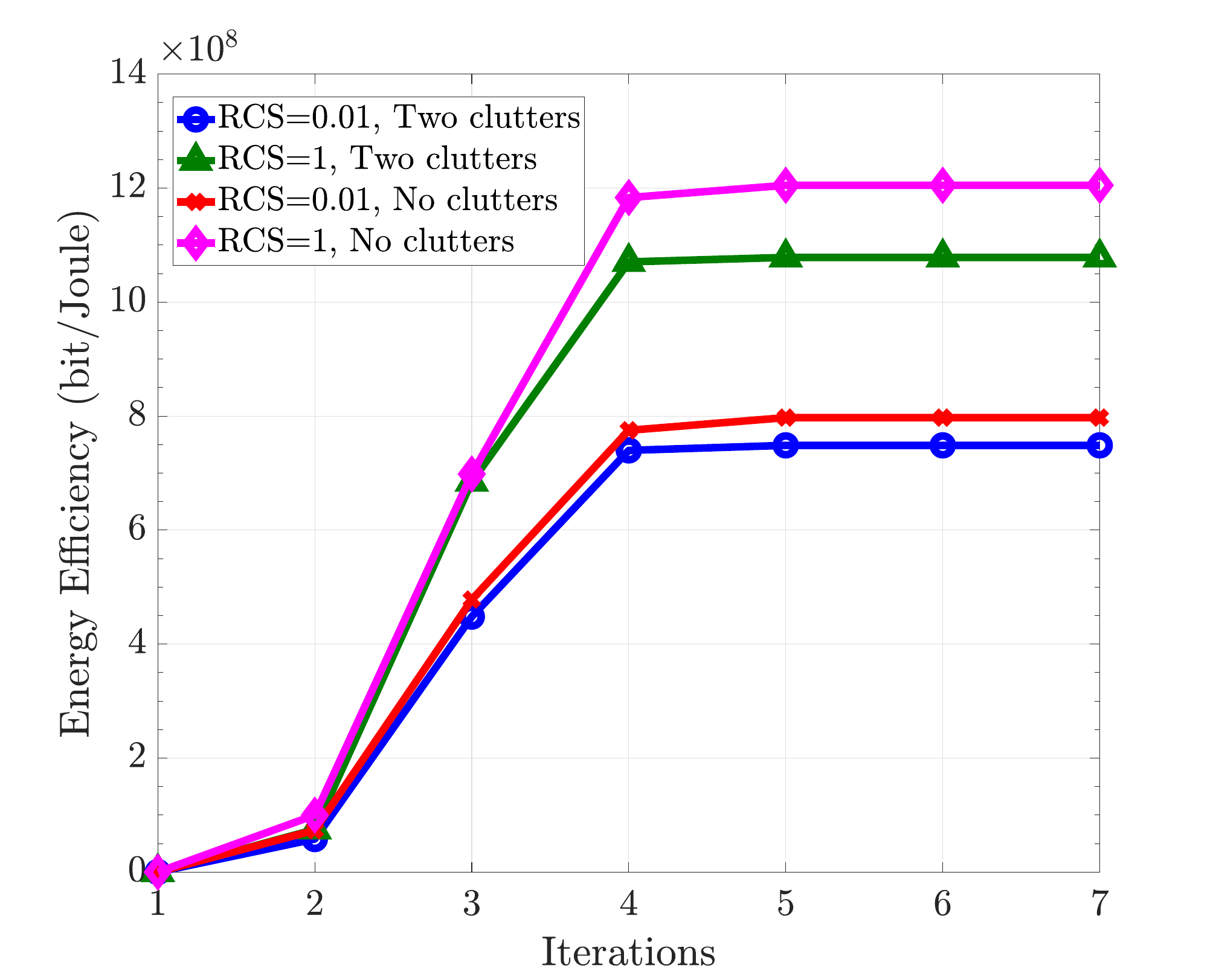}
      \caption{{{Convergence behavior of the proposed Algorithm~\ref{algo1}.}}}
       \label{fMIDRPL}
\end{minipage}\hfill
 \begin{minipage}{0.32\textwidth}
 \centering
 \includegraphics[width=1.05\linewidth]{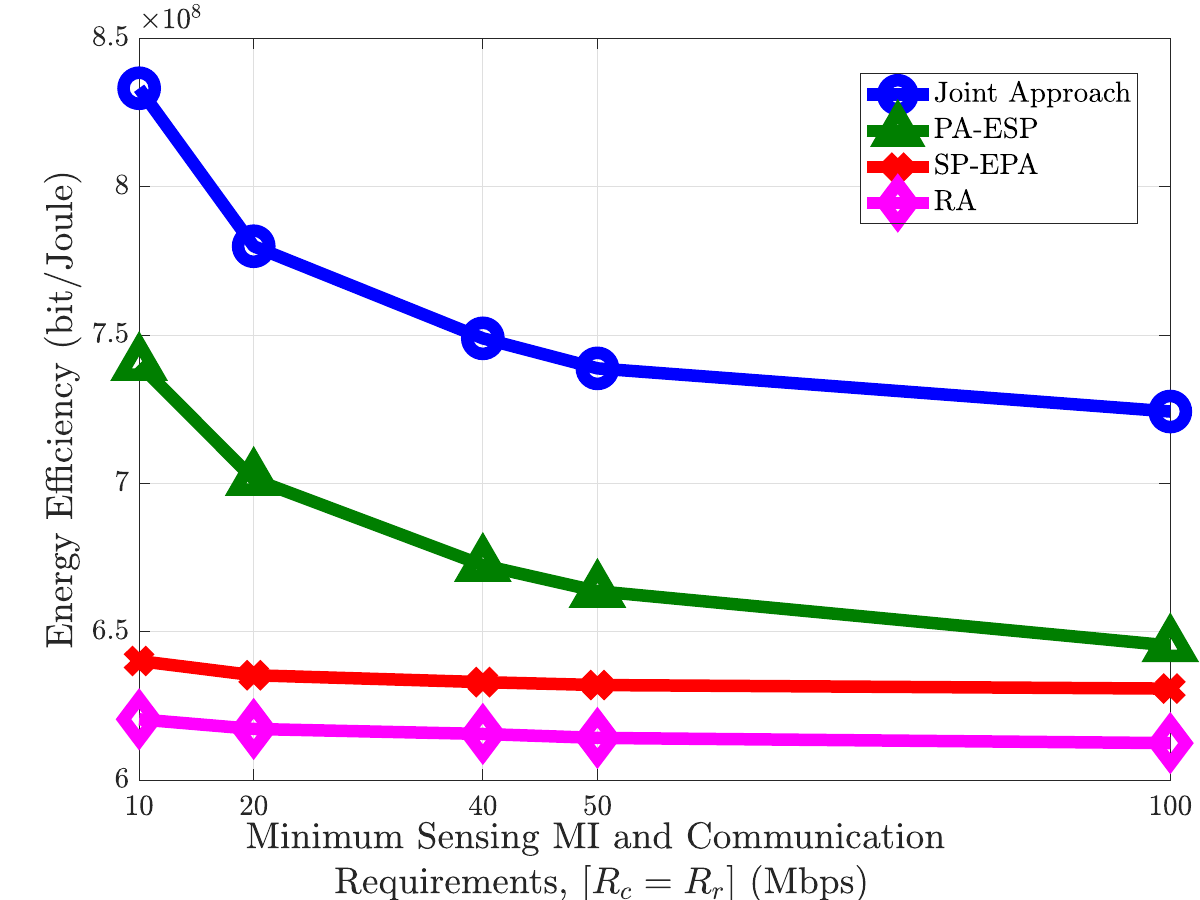}
      \caption{{{EE as a function of sensing MI threshold ($R_r$) and data rate threshold ($R_c$) when $R_c=R_r$.}}}
       \label{EE_vs_QoS}
\end{minipage}%
\vspace*{-3mm}
\end{figure*}

{{Figure~\ref{MI_rcs} depicts the impact of varying the values of the RCS and the maximum transmit power on the sum of MI and data rate for the proposed approach. First, it can be observed that increasing the value of the RCS leads to higher sensing MI and data rate. This is because, the RCS value physically represents the area into which targets reflects incoming radar signal. The greater the RCS, the stronger the reflected radar echoes, leading to higher signal strength for sensing purposes. Second, augmenting the transmit power budget leads to higher aggregate sensing MI and data rate. This observation is the same for different RCS values. The reason is, increasing the value of the maximum transmit power results in a larger feasible region for the joint optimization problem.}}

Figure~\ref{fMIDRPL} illustrates the EE and convergence behavior of Algorithm~\ref{algo1} under different RCS and number of clutters for ISaC and sensing-only users. It can be seen that  the proposed algorithm converges within no more than 5 iterations on average under different RCS and number of clutters, which verifies the fast convergence of Algorithm~\ref{algo1}.
{\textcolor{black}{Finally, Figure~\ref{EE_vs_QoS} compares the average EE of the proposed joint approach and the considered benchmarks for different minimum communication and sensing QoS requirements. It can be observed that the proposed scheme achieves superior EE than the benchmarks, demonstrating the significance of joint SA and PA. The EE reduces with increasing $R_c$ and $R_r$ since higher QoS threshold restricts the feasible region.}}



\vspace*{-2mm}
\section{Conclusion}
In this letter, a unified joint PA and SP optimization framework to maximize the sum of MI and data rate, and EE in a semi-ISaC system has been proposed. The sum of MI and data rate optimization problem was shown to be convex and can be solved using standard convex optimization solvers. On the other hand, the non-convex EE problem was transformed into into a concave-convex fractional problem that can be solved via the Dinkelbach-method. Simulation results have demonstrated the efficiency and the fast convergence of the  proposed joint schemes, the advantage of allocating more spectrum resources to the ISaC user, and the impact of varying user priority on the performance of the semi-ISaC network.
\textcolor{black}{Further research is required to consider the impact of imperfect channel estimation and multi-antenna BSs.}

\vspace*{-3mm}
\bibliographystyle{IEEEtran}
\bibliography{IEEEfull}

\end{document}